\documentclass[a4paper,twocolumn,superscriptaddress]{revtex4}
\usepackage[latin1]{inputenc}
\usepackage{graphicx}
\usepackage{amsfonts}
\usepackage{amsmath}
\usepackage{amssymb}
\usepackage{verbatim} 
\usepackage{latexsym}
\usepackage{indentfirst}
\usepackage{amsmath}
\usepackage{subfigure}
\usepackage{mathrsfs}

\begin{document} 
\title{Hard thermal loops in long wave-length and static external gravitational fields}

\author{R. R. Francisco}
\affiliation{Instituto de Física, Universidade de São Paulo, 05508-090, São Paulo, SP, Brazil}
\author{J. Frenkel} \thanks{e-mail: rarofra@fma.if.usp.br, jfrenkel@fma.if.usp.br}
\affiliation{Instituto de Física, Universidade de São Paulo, 05508-090, São Paulo, SP, Brazil} 

\begin {abstract}

We study, in the long wave-length and static limits, the structure of the $n$-point graviton functions at high temperature. Using the gauge and Weyl invariance of the theory, we derive a simple expression for the hard thermal amplitudes in these two limits.
\end {abstract}

\maketitle

An important issue in thermal pertubation theory is the study of hard thermal loops (HTL), where the external energies and momenta are much less than the temperature $T$ \cite{1}. These loops are relevant in the resummation procedure which is necessary to control the perturbative infra-red divergences. Such thermal amplitudes have simple gauge and symmetry properties, but are in general non-local functionals of the external fields \cite{2,3}. However, in the long wave-length limit (L) where the external fields are position-independent, and in the static limit (S) where the fields are time-independent, the hard thermal amplitudes become local functions of the external fields. It turns out that in these limits, such amplitudes become (in momentum space) independent of the external energies and momenta. Nevertheless, these two limits lead to different functions for the thermal amplitudes \cite{4,5}.

The main purpose of this letter is to derive a simple general expression for the hard thermal loops which arise in the long wave-length and static limits (See Eq.(\ref{Eq.20})). Since in these limits higher point functions vanish in external electromagnetic and Yang-Mills fields \cite{4}, we restrict here to the case of HTL in external gravitational fields. We show, from an analysis of hard thermal perturbation theory, that invariances under gauge and Weyl transformations are sufficient to determine uniquely the structure of the HTL in these limits.

\begin{figure}[t]
\centering
\includegraphics[width=0.45 \textwidth]{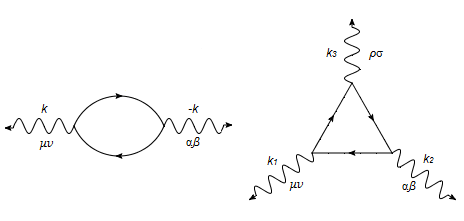}
\caption{One loop thermal diagrams. Solid lines denote thermal particles and wavy lines represent the external gravitational fields.}
\label{FIG.1}
\end{figure}

We shall consider first, as an example, the 3-graviton amplitude, but try to present the argument in a form which makes clear how to extend it to high orders. According to \cite{6,7}, this thermal amplitude involves an angular integration of a Lorentz tensor $C_{\mu \nu , \alpha \beta, \rho \sigma}(Q,k_{1},k_{2},k_{3})$ where $k_{i}$ are the external momenta (FIG.\ref{FIG.1}), $Q_{\mu}=(1,\hat{Q})$ and the angular integration is over the direction of the unit vector $\hat{Q}$.

We represent the gravitational metric as:

\begin{equation}
\label{Eq.1}
\sqrt{-g}g^{\mu \nu}=\eta^{\mu \nu}+h^{\mu \nu} 
\end{equation}
where $\eta^{\mu \nu}$ is the Minkowski metric and expand the perturbative series in powers of $h^{\mu \nu}$. We then find that the Ward identity connecting the 3-graviton and 2-graviton amplitudes leads to the relation:
\begin{multline}
\label{Eq.2}
2k_1^\mu C_{\mu \nu , \alpha \beta, \rho \sigma}(Q,k_{1},k_{2},k_{3})
=[-k_{1\alpha}C_{\nu \beta, \rho \sigma}(Q,k_{3})\\-(\alpha \leftrightarrow \beta) +k_{2\nu}C_{\alpha \beta, \rho \sigma}(Q,k_{3})]
+[(k_{2}, \alpha , \beta) \leftrightarrow (k_{3}, \rho , \sigma)]
\end{multline}

From (\ref{Eq.2}) we deduce that the 3-graviton tensor $C_{\mu \nu , \alpha \beta, \rho \sigma}$ does not have any terms containing the Minkowski metric. The reason is that the 2-graviton tensors on the right hand side of (\ref{Eq.2}) do not contain this metric (See (\ref{Eq.3},\ref{Eq.4}) and, moreover, there is no $k_{1\nu}$ on the right hand side as there would be if the 3-graviton tensor contained $ \eta^{\mu \nu}$. Thus, the Lorentz tensor $C_{\mu \nu , \alpha \beta, \rho \sigma}$ could, in principle, be proportional to some product of the vectors $Q_{\mu},k_{1\mu},k_{2\mu},k_{3\mu}$. However, since in the long wave-length and static limits the HTL are independent of the external momenta, it follows that in these limits the 3-graviton tensor would only involve a product of $Q_{\mu}$.

The above argument may be extended iteratively to higher orders, leading to the conclusion that all graviton tensors arising in hard thermal perturbation theory involve, in the long wave-length and static limits, only a symmetric product of the vector $Q_{\mu}$. For instance, we give here the form of the 2-point graviton functions in these two limits:

\begin{equation}
\label{Eq.3}
\Gamma_{\mu _{1}\nu _{1},\mu _{2}\nu _{2}}=\frac{\mathscr{C} \pi T^{4}}{240}\left ( -\frac{1}{2}\right ) \int d\Omega \; C_{\mu _{1}\nu _{1},\mu _{2}\nu _{2}}(Q)
\end{equation}
where the Casimir $\mathscr{C}$ gives the number of internal degrees of freedom, and:

\begin{subequations}
\label{Eq.4}
\begin{eqnarray}
C^{L}_{\mu _{1}\nu _{1},\mu _{2}\nu _{2}}(Q)=(w-1)Q_{\mu _{1}}Q_{\nu _{1}}Q_{\mu _{2}}Q_{\nu _{2}} \label{Eq.4a}
\\
C^{S}_{\mu _{1}\nu _{1},\mu _{2}\nu _{2}}(Q)=(w-5)Q_{\mu _{1}}Q_{\nu _{1}}Q_{\mu _{2}}Q_{\nu _{2}} \label{Eq.4b}
\end{eqnarray}
where $w$ is the number of Lorentz indices which are equal to zero.
\end{subequations}

We note that the Ward identities demand that loops with $n$ external lines must have the same $T^{4}$ dependence for all $n$. In particular, there is also a tadpole contribution for $n=1$, which is given by:

\begin{equation}
\label{Eq.5}
\Gamma_{\mu_{1}\nu_{1}}=\frac{\mathscr{C} \pi T^{4}}{240}\int d\Omega \; C_{\mu_{1} \nu_{1} }(Q)=\frac{\mathscr{C} \pi T^{4}}{240}\int d\Omega \; Q_{\mu_{1}}Q_{\nu_{1}}
\end{equation}

Using the simple structure we have just deduced for the graviton tensors, we shall next determine all the hard higher point functions. We will proceed by induction, starting from (\ref{Eq.3}) and (\ref{Eq.5}), and assuming that the $n$-point graviton amplitudes in the above limits have the form:

\begin{equation}
\label{Eq.6}
\Gamma_{\mu_{1}\nu_{1},...,\mu_{n}\nu_{n}}=\frac{\mathscr{C} \pi T^{4}}{240}\left ( -\frac{1}{2}\right )^{n-1}\int d\Omega \; C_{\mu _{1}\nu _{1},...,\mu _{n}\nu _{n}}(Q)
\end{equation}
where, for $n\geqslant 2$:
\begin{subequations}
\label{Eq.7}
\begin{multline}
\label{Eq.7a}
C^{L}_{\mu _{1}\nu _{1},...,\mu _{n}\nu _{n}}(Q)=
\\(w-1)...(w-2n+3)Q_{\mu _{1}}Q_{\nu _{1}}...Q_{\mu _{n}}Q_{\nu _{n}} 
\end{multline}
\begin{multline}
\label{Eq.7b}
C^{S}_{\mu _{1}\nu _{1},...,\mu _{n}\nu _{n}}(Q)=
\\(w-5)...(w-2n-1)Q_{\mu _{1}}Q_{\nu _{1}}...Q_{\mu _{n}}Q_{\nu _{n}} 
\end{multline}
\end{subequations}
and $w$ denotes the number of Lorentz indices which have the value $0$.

We now prove that also the $(n+1)$-point graviton amplitudes have the same basic structure as the one shown in (\ref{Eq.6},\ref{Eq.7}). To this end, we use the fact these amplitudes obey the Weyl identity which leads to:
\begin{multline}
\label{Eq.8}
\eta^{\mu_{n+1}\nu_{n+1}}C{\mu _{1}\nu _{1},...,\mu _{n}\nu _{n},\mu _{n+1}\nu _{n+1}}(Q)=
\\-2nC{\mu _{1}\nu _{1},...,\mu _{n}\nu _{n}}(Q)
\end{multline}

In view of our previous discussion, we may write the tensor on the left hand side of (\ref{Eq.8}) in the form:
\begin{multline}
\label{Eq.9}
C{\mu _{1}\nu _{1},...,\mu _{n}\nu _{n},\mu _{n+1}\nu _{n+1}}(Q)=
\\P_{n}(w)Q_{\mu _{1}}Q_{\nu _{1}}...Q_{\mu _{n}}Q_{\nu _{n}}Q_{\mu _{n+1}}Q_{\nu _{n+1}}
\end{multline}
where $P_{n}(w)$ is a polynomial in $w$. With the help of (\ref{Eq.6}) and(\ref{Eq.8}), we then obtain the relations:
\begin{subequations}
\label{Eq.10}
\begin{eqnarray}
P_{n}^{L}(w+2)-P_{n}^{L}(w)=2n(w-1)...(w-2n+3) \label{Eq.10a}
\\ P_{n}^{S}(w+2)-P_{n}^{S}(w)=2n(w-5)...(w-2n-1) \label{Eq.10b}
\end{eqnarray}
\end{subequations}

These two equations show, firstly, that $P_{n}(w)$ must be a polynomial of degree $n$ in $w$. Secondly, these provide $(n-1)$ recurrence relations which fix uniquely the coefficients of this polynomial, up to a constant term which cancels in the differences on the left hand side of (\ref{Eq.10}). But this term may be determined by a futher use of the Ward identity. For example, in the long wave-length limit and when all indices are $0$, this identity leads to the condition:
\begin{equation}
\label{Eq.11}
P_{n}^{L}(w=2n+2)=(2n+1)P_{n-1}^{L}(w=2n)=(2n+1)!!
\end{equation}

This, together with the relation (\ref{Eq.10a}), completely fixes the polynomial $P_{n}^{L}(w)$, leading to a result in accordance with the form assumed in (\ref{Eq.7a}). Proceeding in a similar way, we obtain for $P_{n}^{S}(w)$ a result consistent with that in (\ref{Eq.7b}), which concludes our inductive reasoning. From these equations it follows that the graviton amplitudes vanish when $w$ is odd, which is consistent with the symmetry properties of the angular integral in (\ref{Eq.6}).

The effective actions which generate, in the static and long wave-length limits, the hard thermal loops in external gravitational fields are given by \cite{4,8,9}: 
\begin{equation}
\Gamma^{S}=\frac{\mathscr{C}\pi T^{4}}{90}\int d^{4}x\frac{\sqrt{-g}}{g^{2}_{00}} \label{Eq.12}
\end{equation}
\begin{equation} \Gamma^{L}=-\frac{\mathscr{C}\pi T^{4}}{120}\int d^{4}x\int d\Omega \left[\left (\frac{g^{0i}Q_{i}}{g^{00}}\right )^{2}-\frac{g^{ij}Q_{i}Q_{j}}{g^{00}} \right]^{1/2} \label{Eq.13}
\end{equation}

These actions are invariant under the corresponding gauge transformations as well as under Weyl transformations. But, in view of the apparent differences between the above forms, it may seen a bit surprising that these actions could generate the static and long wave-length amplitudes (\ref{Eq.6},\ref{Eq.7}), which have a rather similar structure.

In order to understand these features, we will use an alternative closed form for the effective actions, which is motivated by a Boltzmann equation approach. In this method, the hard thermal effective action can be written as \cite{10}:

\begin{equation}
\label{Eq.14}
\Gamma=\frac{\mathscr{C}}{(2\pi)^{3}}\int d^{4}x \int d^{4}p \; \theta(p_{0})\theta(g^{\mu\nu}p_{\mu}p_{\nu})N(P)
\end{equation}
where $N$ is the thermal distribution function and $P$ is a constant of motion. Let us consider, for example, the long wave-length limit when $P=\sqrt{p_{i}p_{i}}$. This form arises because, in this case, $p_{i}$ is a solution of the Boltzmann equation, which is invariant under space-independent gauge transformations. Using the metric representation given in (\ref{Eq.1}), and making the change of variables:

\begin{equation}
\label{Eq.15}
p_{i}=\hat{Q_{i}}P; \;\;\;|\hat{Q}|=1; \;\;\;p_{0}=zP
\end{equation}
we may perform the $P$ integration in (\ref{Eq.14}). In this way, the expression giving the effective action in the long wave-length limit becomes:

\begin{multline}
\Gamma^{L}=\frac{\mathscr{C}\pi T^{4}}{120}\int d^{4}x\int d\Omega\int_{0}^{\infty}dz
\\ [\theta(z^{2}-1+Az^{2}+2Bz+C)-\theta(z^{2}-1)] \label{Eq.16} 
\end{multline}
where
\begin{equation}
\label{Eq.17}
A=h^{00}; \;\;\;B=h^{0i}Q_{i}; \;\;\;C=h^{ij}Q_{i}Q_{j}
\end{equation}
and we have normalised $\Gamma^{L}$ so that it vanishes at $h^{\mu\nu}=0$.

We now expand the $\theta$-function in (\ref{Eq.16}) in powers of $h^{\mu\nu}$. Let $z=\sqrt{u}$. Then we get, for the $n$-th term, an integral of the form:

\begin{equation}
\label{Eq.18}
\frac{1}{n!}\int_{0}^{\infty}du\frac{1}{2\sqrt{u}}\delta^{n-1}(u-1)[Au+2B\sqrt{u}+c]^{n}
\end{equation}
where $\delta^{n-1}$ is the $(n-1)$th derivative of the delta function. Let us take out of the expression in the square bracket in (\ref{Eq.18}), a term proportional to $u^{w/2}$, that is, $z^{w}$. For this term, assuming $n\geqslant 2$, we get the contribution:

\begin{multline}
\label{Eq.19}
\frac{1}{2n!}\int_{0}^{\infty}du\;\delta^{(n-1)}(u-1)u^{(w-1)/2}=
\\ \frac{1}{2n!}\left (-\frac{1}{2}\right )^{n-1}(w-1)...(w-2n+3)
\end{multline}

From (\ref{Eq.16}-\ref{Eq.18}) one can see that for a given $n$, the coefficient of $z^{w}$ involves a product of $n$ functions $h^{\mu\nu}$ such that the number of Lorentz indices which have the value $0$ is equal to $w$. This property, together with the result (\ref{Eq.19}), show that the effective action (\ref{Eq.14}) generates in the long wave-length limit, an amplitude which is equivalent to that given in (\ref{Eq.6}) and (\ref{Eq.7a}). Similarly, one can show that in the static limit, the action (\ref{Eq.14}) with $P=p_{0}$, precisely generates the amplitudes given in (\ref{Eq.6}) and (\ref{Eq.7b}).

Thus, we conclude that in the long wave-length and static limits, the hard thermal loops in external gravitational fields may be expressed in the simple form:
\begin{multline}
 \label{Eq.20}
\Gamma_{\mu_{1}\nu_{1},...,\mu_{n}\nu_{n}}=\frac{\mathscr{C} \pi T^{4}}{240}\left ( -\frac{1}{2}\right )^{n-1}
\\P_{n-1}(w)\int d\Omega \; Q_{\mu_{1}}Q_{\nu_{1}}...Q_{\mu_{n}}Q_{\nu_{n}}
\end{multline}
where $Q_{\mu}=(1,\hat{Q})$, $w$ is the number of time-like Lorentz indices and $P_{n-1}(w)$ are polynomials of degree $(n-1)$ in $w$. These are equal to $1$ when $n=1$ and for $n\geqslant 2$ they are given by:
\begin{subequations}
\label{Eq.21}
 \begin{eqnarray}
  P^{L}_{n-1}(w)=(w-1)...(w-2n+3)\label{Eq.21a}
\\P^{S}_{n-1}(w)=(w-5)...(w-2n-1)\label{Eq.21b}
 \end{eqnarray}
\end{subequations}

We note that the angular integral in (\ref{Eq.20}) vanishes when $w$ is odd. When $w$ is even, it is straightforward to evaluate this integral in terms of products of Kronecker delta functions, but the result is somewhat cumbersome to write down. The above amplitudes lead (when summed to all orders) to the static and long wave-length effective actions (\ref{Eq.12}) and (\ref{Eq.13}), which are local functions of the external gravitational fields.
\\

We would like to thank FAPESP and CNPq (Brazil) for a grant. J. F. is indebted to Prof. J. C. Taylor for a very helpful correspondence.


\begin {thebibliography}{10}
\bibitem{1} M. Le Bellac, Thermal Field Theory, Cambridge Univ. Press, Cambridge, 2000
\bibitem{2} J. Frenkel, J. C. Taylor, Nucl. Phys. B 334 (1990) 199; Nucl. Phys. B 374 (1992) 156
\bibitem{3} E. Braaten, R. D. Pisarski, Nucl. Phys. B 337 (1990) 569; nucl. Phys. B 339 (1990) 310
\bibitem{4} F. T. Brandt, J. Frenkel, J. C. Taylor, Nucl. Phys. B 814 (2009) 366
\bibitem{5} J. Frenkel, S.H. Pereira, N. Takahashi, Phys. Rev. D 79 (2009) 085001
\bibitem{6} F. T. Brandt, J. Frenkel, J. C. Taylor, Nucl, Phys. B 374 (1992) 169
\bibitem{7} F. T. Brandt, J. Frenkel, Phys. Rev. D 47 (1993) 4688
\bibitem{8} A. Rebhan, Nucl. Phys. B 351 (1991) 706
\bibitem{9} J. Frenkel, J. C. Taylor, Z. Phys. C 49 (1991) 515
\bibitem{10} F. T. Brandt, J. Frenkel, J. C. Taylor, Nucl. Phys.  437 (1995) 433
\end {thebibliography}
\end{document}